\documentstyle[12pt,epsf]{article}

\newcommand{\bce}{\begin{center}}
\newcommand{\ece}{\end{center}}
\newcommand{\bea}{\begin{eqnarray}}
\newcommand{\eea}{\end{eqnarray}}
\newcommand{\be}{\begin{equation}}
\newcommand{\ee}{\end{equation}}
\newcommand{\bd}{\begin{displaymath}}
\newcommand{\ed}{\end{displaymath}}
\newcommand{\bit}{\begin{itemize}}
\newcommand{\eit}{\end{itemize}}
\newcommand{\ben}{\begin{enumerate}}
\newcommand{\een}{\end{enumerate}}
\newcommand{\bdes}{\begin{description}}
\newcommand{\edes}{\end{description}}
\newcommand{\wt}{\mbox{$\rm W_t$}}
\newcommand{\wtn}{\mbox{${\rm W}_{\rm t}^{(n)}$}}
\newcommand{\Wt}[1]{\mbox{${\rm W}_{\rm t}^{(#1)}$}}
\newcommand{\E}{\> = \>}
\newcommand{\EA}{&=&}
\newcommand{\bfl}{\begin{flushright}}
\newcommand{\efl}{\end{flushright}}

\begin{document}

\thispagestyle{empty}
\vspace{2cm}

\bce

{\Large\bf The $W_t$ Transcendental Function and 
           Quantum Mechanical Applications} 

\vspace{1cm}

V.~E.~Markushin $^1$, R.~Rosenfelder $^1$ and A.~W.~Schreiber $^2$

\vspace{0.5cm}
$^1$ Paul Scherrer Institute, CH-5232 Villigen PSI, Switzerland\\
$^2$ Department of Physics and Mathematical Physics and
              Research Centre for the Subatomic Structure of Matter,
              University of Adelaide, Adelaide, S. A. 5005, Australia

\ece
\vspace{4cm}
\begin{abstract}
\noindent
We discuss the function $\wt(x)$  defined 
via the implicit equation $ \wt(x) \cdot \tan \left [ \wt(x) \right ] = x $, 
which appears in certain quantum mechanical and field theoretic
applications.  We investigate its analytic structure, develop series 
expansions for both small and large $x$, and provide various techniques for 
its numerical evaluation in the complex plane. 

\end{abstract}

\vspace{7cm}
PACS-numbers: 02.30Gp, 02.30.Mv, 03.65.Db

\newpage

\section{Introduction}

Implicit equations for functions appear frequently in physics although they 
are
not discussed at all in standard handbooks on (standard) special functions 
\cite{Handbook,GR}.
The most famous example, probably, is {\it Kepler's} equation 
\be
\omega t \E \psi \> - \> e \> \sin \psi
\ee
which describes the dependence of the angle $\psi$ 
(``the eccentric anomaly'') 
on the time $t$ for an elliptical orbit with eccentricity $e$ \cite{Gold}. 
It has been
a challenge to solve this equation for generations of astronomers 
and mathematicians,  among them F.~W.~Bessel who was led by this problem 
to the functions which bear his name. 

Another transcendental equation which has made its appearance in a variety of
applications is
\be
{\rm W}(x) \, \exp \left [ {\rm W}(x) \right ] \E x \> ,
\label{Wexp}
\ee
the history of which goes back to J. H. Lambert and L. Euler in the 18th 
century. A comprehensive survey of the many available results for this 
function 
has been given in Ref. \cite{CGHJK} where it has been called the 
{\it Lambert} W-function
(probably because Euler already has enough named relations and functions to 
his credit \footnote{In Ref. \cite{Bel} a slightly more general equation is 
called ``Ramanujan's equation''.}...). Apart from the examples given in the 
above article 
there is also some recent interest in the properties of W$(x)$ in 
particle \cite{Sh}  and nuclear physics \cite {Ro}.

Here we want to study a function defined by a superficially similar equation 
as given in (\ref{Wexp}), viz.
\be
\wt (x) \, \cdot \tan \left [ \wt (x) \right ] \E x
\label{Wt}
\ee
which, however, has a much richer structure than Lambert's W-function. 
Because of the similarity we propose to call 
\footnote{``Names are important'' \cite{CGHJK} for a consistent nomenclature. 
In the present scheme the Lambert W-function
 would be denoted by $ {\rm W}_e(x) $.} the solution of Eq. (\ref{Wt})   
$\wt (x)$. Analogously, one could define
\be
{\rm W_{ct}} (x) \,\cdot  \cot \left [ {\rm W_{ct}} (x) \right ] \E x
\label{Wcot}
\ee 
and ${\rm W}_s(x)$ and ${\rm W}_c(x)$. Eq. (\ref{Wt}) emerged in a variational 
approach to 
Quantum Electrodynamics \cite{ARS} where the relation between the bare mass 
and the physical mass of the electron was obtained in terms of the solution 
of an implicit equation of the type (\ref{Wt}). However, as we will see in 
Section 2, 
the same equation also appears as an eigenvalue equation in a one-dimensional 
quantum mechanical problem for
an infinite square well with a residual $\delta$-function interaction. If the 
strength
of this additional interaction is constant one is naturally led to 
Eq. (\ref{Wcot}) 
whereas a linear energy dependence of this strength leads to Eq. (\ref{Wt}).

In the present, mostly elementary, note we will concentrate on the 
\wt-function and, 
since we are not aware of any other systematic study, we will derive some
representations by simple methods and study the
interesting analytical structure of this function in the complex plane. We 
hope that 
this collection of results will be helpful for other applications where this  
function makes its appearance.

\vspace{1cm}

\section{A quantum mechanical problem leading to $\wt(x)$}
\label{sec:QM}
\setcounter{equation}{0}

Consider a nonrelativistic particle moving in an one-dimensional infinite 
square well in $ 0 < \xi < a $. Assume that an attractive  $\delta$-function 
interaction is added in the middle of the interval whose strength is growing 
with the energy of the particle \footnote{Energy-dependent potentials are frequently
considered in nuclear and hadronic physics where they arise from eliminating
part of the physical Hilbert space.}
\be
\Delta V \E - \lambda \, E \, \delta \left ( \xi - \frac{a}{2} \right ) \> .
\label{delta}
\ee
As the total Hamiltonian is symmetric with respect to transformations 
$\xi \to a - \xi $
the wave functions can be classified according to their symmetry under this 
transformation $ \psi_n (\xi) = \pm \psi_n (a-\xi) $.
Only the even wave functions are influenced by the additional interaction 
(\ref{delta}) since the odd wave functions vanish at $ \xi = a/2 $.  
The Schr\"odinger equation
\be
 - \frac{\hbar^2}{2m} \, \frac{d^2 \psi(\xi) }{d\xi^2} - \lambda \, E \, 
\delta \left ( \xi - \frac{a}{2} \right ) \, \psi(\xi) \E E \psi(\xi) 
\label{Schroedinger}
\ee
with $ E = \hbar^2 k^2/(2m) $ has the solutions
\bea
   \psi(\xi) \EA 
   \left\{  
     \begin{array}{lcl} 
          A_{I}  \sin ( k \xi )        & , & 0 \le \xi <  a/2  \\ 
          A_{II} \sin ( k ( a - \xi) ) & , & a/2 < \xi \le  a  \> .\\ 
     \end{array} 
   \right. 
\label{psi}
\eea
The solution (\ref{psi}) fulfills the boundary conditions that the wave 
function has to vanish at the edges ($ \xi = 0 $ and $\xi = a$ ) of the 
infinite square well. 
At $ \xi = a/2 $ the wave function has to be continuous
\be
 A_I \sin \left ( \frac{k a}{2} \right ) \E   A_{II} \sin \left ( 
\frac{k a}{2} \right) \quad ,
\label{boundary 1} 
\ee
but its derivative makes a jump proportional to the strength of the 
$\delta$-function.
This can be derived in the standard way by integrating Eq. 
(\ref{Schroedinger}) from 
$ a/2 - \epsilon$ to $a/2 + \epsilon$ and letting $\epsilon \to 0$
\be
- \frac{\hbar^2}{2m} \left [ \, \frac{d\psi}{d\xi} \left  (\frac{a}{2} + 
\epsilon \right ) - 
\frac{d\psi}{d\xi} \left ( \frac{a}{2} - \epsilon \right ) \, \right ] - 
\lambda \, E \, \psi  \left ( \frac{a}{2} \right ) \E 0 \> .
\label{jump}
\ee
The r.h.s. vanishes since the wave function is finite and therefore its 
integral over an infinitesimal distance does not give a contribution. For 
our solutions this implies
\be
A_{II} \, k \, \cos \left ( \frac{k a}{2} \right )  + A_I \, k \, \cos 
\left ( \frac{k a}{2}\right ) \E 
\frac{2m}{\hbar^2} \lambda  \, E \,  A_I \sin \left (  \frac{k a}{2} \right )
\E k^2  \lambda \,  A_I \sin \left (  \frac{k a}{2} \right ) \> .
\label{boundary 2}
\ee
Combining the boundary conditions (\ref{boundary 1}) and (\ref{boundary 2}) 
we obtain as
eigenvalue condition \footnote{Note that the energy-dependence of the 
additional interaction is essential; without that one would obtain the 
relation 
$ (k a/2) \cot (k a/2) = \lambda a m /(2 \hbar^2) $ leading to the 
${\rm W}_{ct}$-function.}
\be 
\frac{k a}{2} \, \tan  \left (  \frac{k a}{2} \right ) \E \frac{a}{\lambda} 
\label{eigenvalue eq}
\ee
with the general solution 
\be
k \E \frac{2}{a} \, \wt \left ( \frac{a}{\lambda} \right ) \> .
\label{wavenumber by wt}
\ee

\begin{figure}[htb]
\bce
\mbox{\epsfysize=10cm\epsffile{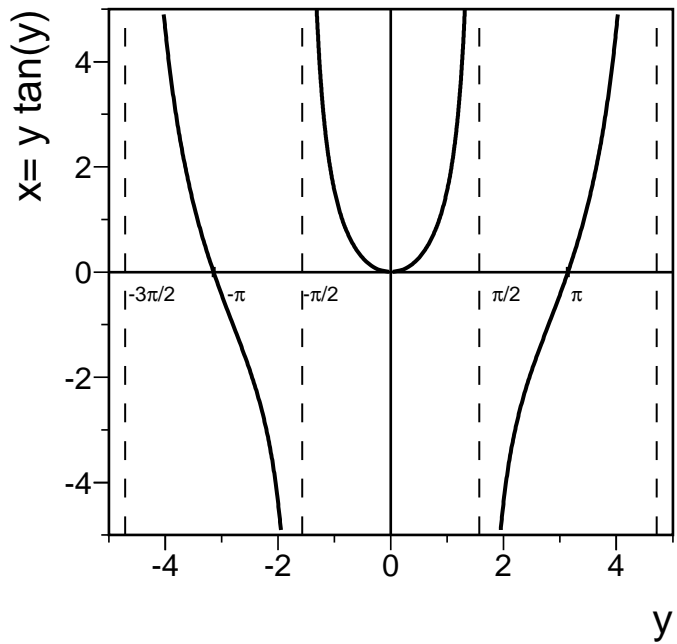}}   
\ece
\vspace*{-1.2cm}
\caption{Eigenvalue equation (\protect{\ref{eigenvalue eq}}) with 
$x = a/\lambda$ and $y = k a/2 $.}
\label{fig:yty}
\vspace{0.5cm}
\end{figure} 

\begin{figure}[htb]
\bce
\mbox{\epsfysize=10cm\epsffile{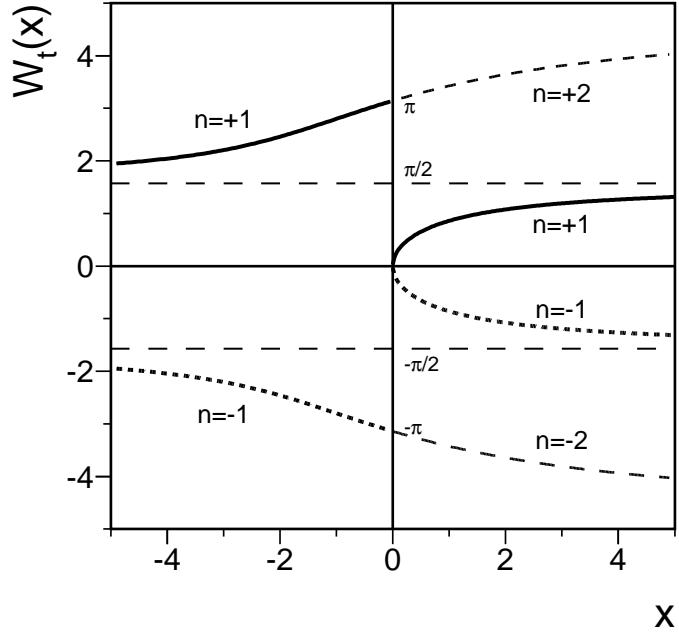}}  
\ece
\vspace*{-1.4cm}
\caption{Branches of the \wt-function.}
\label{fig:branches}
\vspace{0.2cm}
\end{figure}

The graphical representation of the eigenvalue equation in
Fig. \ref{fig:yty} shows that \wt\ is a {\it multivalued} function, and
we have to define its branches. Our definition will be motivated by
physical arguments: for vanishing interaction ($\lambda = 0$) the r.h.s.
of Eq. (\ref{eigenvalue eq}) diverges and thus the
$\tan$-function has to diverge as well. Therefore 
\be
 \frac{k a}{2} \bigg|_{\lambda = 0} \E  \pm \frac{\pi}{2}, \ \pm  
\frac{3\pi}{2},\ \ldots   
\ee 
which obviously gives the unperturbed
wavenumbers (and hence energies) of the symmetric states in the
infinitely deep square well of length $a$. Thus we define the $n$th 
branch of \wt\ by the
requirement that $x = \infty$ is a regular point and 
\be
  \wtn(x \to \infty) \E  {\mathrm sgn}(n) \, \left ( |n| - \frac{1}{2} 
\right ) \, \pi \quad ,  \quad n \E \pm 1,\ \pm 2,\ \ldots
\label{branch}
\ee
The limit of very strong coupling ($\lambda \to \infty$) is also easily 
understood: 
if the $\delta$-function is attractive the energy levels are lowered:
\be
  \frac{k a}{2}\bigg |_{\lambda \to +\infty} \E  0, \ \pm \pi, \ \ldots \> ,
\ee
i.e.
\be
  \wtn(x \to +0) \E  {\mathrm sgn}(n) \, \left ( |n| - 1 \right ) \, 
\pi \quad ,  \quad n \E \pm 1,\ \pm 2,\ \ldots  \> .
\label{x to +0}
\ee
However, if the $\delta$-function is very much repulsive the square well is 
practically halved and the eigenvalue condition is $\sin(k a/2) = 0 $. 
This is formally the same as for $ \lambda \to + \infty$ but
$ k = 0 $ is excluded as trivial eigenvalue. Therefore
the levels are pushed up:
\be
  \frac{k a}{2}\bigg |_{\lambda \to -\infty} \E   \pm \pi, \ \pm 2 \pi , \ 
\ldots
\ee
i.e.
\be
  \wtn(x \to -0) \E n \, \pi \>  \> .
\label{x to -0}
\ee
These conventions are summarized in Fig.~\ref{fig:branches}.  
Since the value of \wt(0) depends on the way one approaches the point 
$ x = 0 $, 
it must be a {\it branchpoint} for the function. Indeed, replacing for 
small argument
the $\tan$-function by its argument one finds for the principal branch 
(which corresponds to the ground state of the system) 
\be
  \Wt{1}(x) \> \stackrel{x \to +0}{\longrightarrow} \> \sqrt{x} \> .
\ee
In general we have from eqs. (\ref{x to +0}, \ref{x to -0})
\be
   \Wt{n}(-0) \E  \Wt{n + {\rm sgn}(n) }(+0)  
\ee
and our branch conventions can be summarized by writing the defining 
Eq. (\ref{Wt}) in the form
\bea
  \wtn(x) \EA  {\rm sgn}(n) \, \left ( |n| - \frac{1}{2} \right ) \, 
\pi + \frac{1}{2 i} 
           \ln \left [ \, \frac{ x - i \wtn(x)}{x + i \wtn(x)} \, \right ] \\
\EA  {\rm sgn}(n) \left ( |n| - \frac{1}{2} \right ) \pi + \Theta(-x) 
\, {\rm sgn} \left [ \wtn(x) \right ]  \pi + 
{\rm arg} \left [ x - i \wtn(x) \right ] 
\label{branch conv}
\eea
where $\ln(z)$ denotes the {\it principal} branch of the logarithmic function
($ - \pi < {\rm arg}(z) \le \pi$). 

\noindent
From Eq. (\ref{branch conv}) we obtain the important relation 
\bea
\Wt{-n}(x) \! \EA  \! - {\rm sgn} (n) \left ( |n| - \frac{1}{2} \right ) \pi +
\Theta(-x) \, 
 {\rm sgn} \left [ \, \Wt{-n}(x) \, \right ]  \pi + 
{\rm arg} \left [ \, x - i \Wt{-n} (x) \, \right ] \nonumber \\
          \EA  -  {\rm sgn} (n) \left ( |n| - \frac{1}{2} \right ) \pi  
- \Theta(-x) \, 
 {\rm sgn} \left [ \, - \Wt{-n}(x) \, \right ]  \pi -
{\rm arg} \left [ \, x + i \Wt{-n} (x) \, \right ] \nonumber \\
\EA  - \Wt{n}(x)
\label{negative n} 
\eea
which allows us to restrict our analysis to branches with $ n > 0 $ 
(cf. Fig. \ref{fig:branches}).
It is, of course, easy to find some special values of the \wt-function, e.g.
\be
\Wt{n+1} \left ( \frac{\pi}{4}+ n \pi  \right ) \E  \frac{\pi}{4} + n \pi \> 
, \quad n \E 0, \ 1, \ \ldots
\label{special}
\ee
The above  definition of the branches of $\Wt{n}(x)$ is still incomplete 
because we have not yet specified  how different sheets of the Riemann 
surface are connected to each other across the cuts that must also be 
explicitly defined.  
This will be discussed in more detail in Section~\ref{SecAStr}.  In the next 
two Sections, our discussion concerns only the properties of the 
main branch $\Wt{1}(x)$ 
on the real $x$ axis, where $\Wt{1}(x)$ is continuous except at the point 
$x=0$.

\section{Series expansions, differential equation and integrals}
\setcounter{equation}{0}

\begin{figure}[htb]
\bce
\mbox{\epsfysize=10cm\epsffile{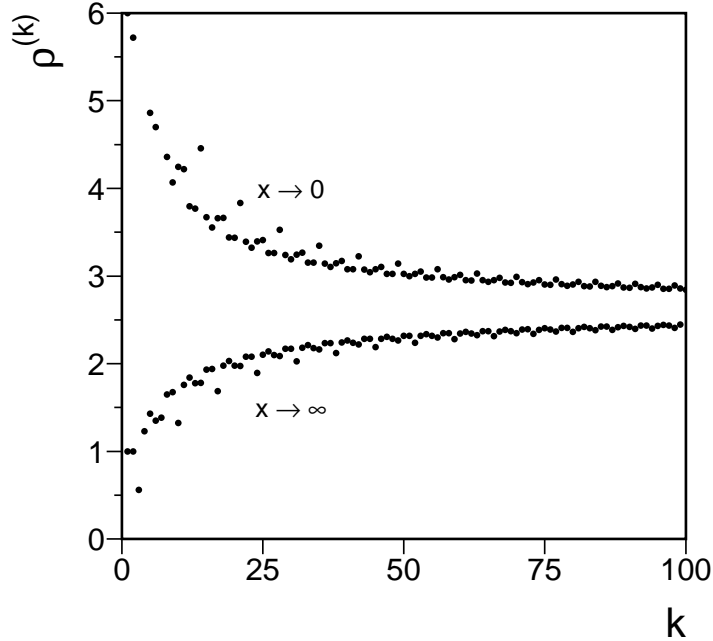}}   
\ece
\vspace*{-1.2cm}
\caption{Estimates of the convergence radius for the series 
expansions (\protect{\ref{small x}}, \protect{\ref{large x}}) of $\Wt{1}(x)$ 
from the first 100 coefficients.}
\label{fig:conv}
\vspace{0.5cm} 
\end{figure}

In this Section we concentrate on the the principal branch of the 
\wt-function, i.e. $ n = 1 $.
From the defining equation (\ref{Wt}) it is easy to obtain the first terms 
in the series expansion for small $x \ge  0$ 
\be
   \Wt{1}(x) \E \sqrt{x} \sum_{k=0}^{\infty} a_k \, x^k 
              \E \sqrt{x} \, \left [ \, 
                1  - \frac{1}{6} x + \frac{11}{360} x^2 - \frac{17}{5040} x^3 
                     - \frac{281}{604800} x^4 + \ldots \, \right ]
\label{small x}
\ee
and, by setting $\wt = \pi (1-v)/2 $ for large $|x|$, 
\be
    \Wt{1}(x) \E \frac{\pi}{2} \sum_{k=0}^{\infty}  \frac{b_k}{x^k} 
             \E  \frac{\pi}{2} \left [ \, 
                 1 - \frac{1}{x} + \frac{1}{x^2} 
                - \left ( 1 - \frac{\pi^2}{12} \right )  \frac{1}{x^3} 
             - \left ( \frac{\pi^2}{3} - 1 \right ) \frac{1}{x^4} + \ldots \, 
\right ].
\label{large x}
\ee
To calculate the expansion coefficients for $ |x| \to \infty$ 
systematically, one may use Lagrange's expansion theorem for implicit 
functions \footnote{See, e.g., Ref. \cite{Handbook}, Eq. 3.6.5 .}:
\be
b_0 \E 1 \> , \hspace{0.3cm} b_k \E - \left ( 
\frac{\pi}{2} \right )^k \frac{1}{k !} \, \frac{d^{k-1}}{d v^{k-1}} \, 
\left ( 
\frac{v(1-v)}{\tan(\pi v/2)} \right )^k \Biggr |_{v=0} \> \>, \> \> k \ge 1
 \> .
\ee
The convergence radii of the above two expansions may be estimated 
from the {\it root} test
\be
x \> \le \> \rho_0 \E \lim_{k \to \infty} \left | a_k \right |^{-1/k} \> 
\equiv \> \lim_{k \to \infty} \rho_0^{(k)} \> \> , 
\hspace{1cm}
|x| \> \ge \> \rho_{\infty}  \E \lim_{k \to \infty} \left | b_k 
\right |^{1/k} 
 \> \equiv \lim_{k \to \infty} \rho_{\infty} ^{(k)}\> \> .
\ee
Fig. \ref{fig:conv} shows the result from the first 100 coefficients; 
it seems that these expansions
converge only for finite values of $x < 2.8 $ and $1/|x| < 2.4 $ which 
hints at additional singular points in the complex $x$-plane. We will
locate these {\it branchpoints} of the \wt-function in Section 5.

\unitlength1mm

At high orders in the above expansions the computational cost of 
using Lagrange's expansion soon becomes prohibitive.
Indeed, the data in Fig.~\ref{fig:conv} were generated by developing recursion
relations for the expansion coefficients in Eqs.~(\ref{small x}, 
\ref{large x}).
By differentiating Eq. (\ref{Wt}) with respect to $x$ and eliminating the 
trigonometric functions one arrives at the simple differential equation
\be
\wt'(x) \E \frac{\wt(x)}{x + x^2 + \wt^2(x)} \> .
\label{diff eq}
\ee
Turning first to the expansion around  $|x| \to \infty$, we set $ t = 1/x$ 
and rewrite  the above differential equation as
\be
(1 + t) \frac{d}{d t} \left ( \frac{1}{\wt(1/t)} \right ) - t^2 
\frac{d \wt(1/t)}{d t} \E  \frac{1}{\wt(1/t)} \>.
\ee
With the ansatz 
\be
\frac{1}{\wt(1/t)} \E \frac{2}{\pi} \sum_{k=0}^{\infty} c_k \, t^k
\ee
one obtains a system of coupled recursion relations for the expansion
coefficients $b_k, c_k $
\bea
(k+1) \, c_{k+1} + (k-1) \, c_k \EA \frac{\pi^2}{4} \left ( 1- \delta_{k 0} 
\right ) \, (k -1 ) \, b_{k-1} 
\label{recurs 1}\\
\sum_{j=0}^{k} b_{k-j} \, c_j \EA  \delta_{k 0} \> ,
\label{recurs 2}
\eea
which can be solved successively for $ k = 0, 1, \ldots $ starting with 
$b_0 = 1$.
For asymptotically large values of $k$, we find (numerically) that the 
coefficients $b_k$ approach the simple functional form
\be
b_k \stackrel{k \to \infty}{\longrightarrow}
 {c \over k^{3/2}}\> \left(\rho_\infty\right)^k \> \sin (a k + b)\;\;\;,
\ee
with
$c \simeq 0.584$, $a \simeq 2.25$, $b \simeq -3.59$ and the radius of
convergence of the expansion in $t$ is
$1/\rho_\infty \simeq 1/2.64 \> $ \footnote{In the quantum electrodynamical 
context where the $\wt$-function appeared \cite{ARS}, this implies a 
{\it finite} convergence
radius of the perturbative expansion contrary to the usual 
semi-heuristic arguments \cite{large order} for an essential singularity at 
zero coupling constant.}. 
Indeed, by writing the trigonometric function
as a sum of two exponentials we find that we can absorb the rapid
oscillations in $b_k$ into the definition of a new variable $x'$ 
(or, equivalently, $t'$), obtained by rotating $x$ to $x e^{i a}$ 
(or $x e^{- i a}$ for the second of the two exponentials).  Hence
we anticipate that the branchpoints responsible for the finite radius
of convergence of  $\wt$  are in fact located at $x = \rho_\infty 
e^{\pm i a} \simeq  2.64 \> e^{\pm 0.72 \pi \, i} $, i.e. in the second and
third quadrants of the complex $x$ plane.  Furthermore note that, generically,
the expansion coefficients of the root $(1 - t \rho_\infty)^\kappa$ 
behave, at large orders $k$, like $ \left(\rho_\infty\right)^k/
k^{\kappa+1}$, which leads us to suspect that the branch cuts are of
square root type.  These suspicions will be confirmed 
in Section 5.

For $x \to +0$ on the other hand we set $\Wt{1}(x) = \sqrt{x} \, f(x) $ 
and obtain the 
differential equation
\be 
(x - 1) f(x) + 2 x (x + 1) f'(x) + f^3(x) + \frac{2}{3} x \frac{d}{d x} 
\left ( f^3(x) \right ) \E 0 \> .
\label{low x diff eq}
\ee
The ansatz
\be 
f^3(x) \E  \sum_{k=0}^{\infty}  d_k \, x^k
\ee
now leads to the following system of recursion relations for the coefficients 
\bea
a_k + \left ( 1 - \delta_{k0} \right ) a_{k-1} + 
\frac{1}{3} \, \frac{2 k + 3}{2 k - 1} \, d_k \EA  0 
\label{recur1}\\
{1 \over k}\sum_{j=1}^k \left ( 4 j - k\right ) a_j \, d_{k-j} 
\EA d_k \> , \hspace{0.5cm} k = 1, 2, \ldots
\label{recur2}
\eea
where $a_0 = d_0 = 1$ is the starting point. Eq. (\ref{recur2}) follows from 
the identity $ f \cdot (f^3)' = 3 f^3 \cdot f'$. For large $k$ the 
recurrence relations yield
\be
a_k \> \stackrel{k \to \infty}{\longrightarrow} \> 
 {c \over k^{3/2}}\> \left(\rho_0\right)^{-k} \> \sin (a k + b)\;\;\;,
\ee
with
$c \simeq 0.564$, $b \simeq -3.14$, $\rho_0 = \rho_\infty$ and $a$ 
the same as above.
\vspace{0.1cm}

Finally, we note that many integrals over the \wt-function may be performed
analytically by a partial integration and 
changing variables to $x = y \tan y$:
\be
\int dx \, f \left [ \wt(x) \right ]  \E x  f \left [ \wt(x) \right ] 
\> - \> \int dy \, y\, \tan y \, f'(y)\;\;\;.
\ee
For example,
\bea
\int dx \, \ln \left [ \, \wt(x) \, \right ] \EA  
x \ln  \left [ \, \wt(x) \, \right ] + \ln \, \biggl |  \cos (\wt(x)) \biggr | 
\label{int 1}\\
\int dx \, \ln \left [ \, \sin \left ( \wt(x) \right )  \, \right ] 
\EA  x \ln \left [ \, \sin \left ( \wt(x) \right ) \, \right ] - \frac{1}{2} 
\wt^2(x) 
\label{int 2}\;\;\;,
\eea
and so on.

\noindent
From the latter equation and the asymptotic behaviour of the \wt-function
one immediately obtains the definite integral
\be
\int_0^{\infty} dx \, \ln \left [ \, \sin (\Wt{1}(x)) \, \right ] \E - 
\frac{\pi^2}{8} \> .
\ee
In addition one finds
\be
\int_0^{\pi/4} dx \, \Wt{1}(x) \E \frac{\pi^2}{16} + \frac{\pi}{8} \ln 2 - 
\frac{1}{2} G
\ee
where $ G = 0.91596594 ...$ is Catalan's constant. This is just the special 
case when Young's inequality (Ref. \cite{GR}, Eq. 12.315) applied to 
the integral
\be
\int_0^a dx \, \Wt{1}(x) \> \ge \> a \frac{\pi}{4} +  \frac{\pi}{8} \ln 2 
- \frac{1}{2} G \> , \> a \le \frac{\pi}{4}
\ee
becomes an equality due to the relation (\ref{special}).

\vspace{1.5cm}

\section{Numerical evaluation and approximations to the $\wt(x)$-function}
\setcounter{equation}{0}

\vspace{0.5cm}
\subsection{Chebyshev approximation}

The numerical evaluation of  the \wt-function for real arguments 
is most easily accomplished 
by expanding in  Chebyshev polynomials \cite{NumRec}
\be
\Wt{1}(x) \E \left \{ \begin{array}{l@{\quad , \quad}l}
 \sqrt{x} \sum_{k=0} \alpha_k \, T_k \left ( \frac{2 x}{a} 
- 1 \right ) & 0 \le x \le a \nonumber \\
 \frac{\pi}{2} \sum_{k=0} \beta_k \, T_k \left ( \frac{a}{x}  
\right ) & |x| > a \label{Cheby} \\
\pi\sum_{k=0} \gamma_k \, T_k \left ( \frac{2 x}{a} +1 
\right ) &
-a \le x < 0 \nonumber \> \> .
\end{array} \right.
\ee

\begin{table}[htb]
\bce 
\vspace{0.4cm}
\begin{tabular}{|r|r|r|r|} \hline
     &                        &                       &                 \\ 
$k$  & $\alpha_k$ \quad \quad & $\beta_k$ \quad \quad & $\gamma_k$ \quad  
\quad \\
     &                        &                       &                 \\ 
\hline
     &                        &                       &                 \\
  0  &     0.80600536         &  1.03465858           &  0.82312766     \\
  1  &    -0.16766125         & -0.28291110           &  0.16771494     \\
  2  &     0.02302848         &  0.03258714           &  0.01423939      \\
  3  &    -0.00298934         &  0.00177957           & -0.00442520      \\
  4  &     0.00030980         & -0.00206359           & -0.00095545      \\
  5  &    -0.00001275         &  0.00044900           &  0.00024196      \\
  6  &    -0.00000478         &  0.00002021           &  0.00007926      \\
  7  &     0.00000178         & -0.00004188           & -0.00001706      \\
  8  &    -0.00000038         &  0.00001127           & -0.00000743      \\
  9  &     0.00000006         &  0.00000018           &  0.00000136      \\
 10  &    -0.00000001         & -0.00000111           &  0.00000075      \\
 11  &     0.00000000         &  0.00000035           & -0.00000012      \\
 12  &     0.00000000         & -0.00000001           & -0.00000008      \\
 13  &     0.00000000         & -0.00000003           &  0.00000001      \\
 14  &     0.00000000         &  0.00000001           &  0.00000001      \\
     &                        &                       &                   \\ 
\hline
\end{tabular}
\ece
\caption{Chebyshev expansion coefficients of $\Wt{1}(x)$ 
in Eq. (\protect{\ref{Cheby}}) for $a = 3.5$.}
\label{tab:cheby}
\vspace{0.5cm}
\end{table}

\noindent
Table \ref{tab:cheby} gives the corresponding coefficients for $a =
3.5$, which was found to be a convenient value to use in order to 
divide the different
regions. One should keep in mind that $|T_n(x)| \le 1 $ for $x \in
[-1,+1] $, so that (assuming a rapid decrease of the coefficients) the
absolute value of the last coefficient gives an estimate of the
accuracy with which $\Wt{1}(x)$ has been approximated by the truncated
Chebyshev expansion.

\subsection{Approximation via iteration}

To calculate the $\wt$-function for arbitrary (complex) $x$,
one may naturally try to solve Eq. (\ref{Wt}) by iteration.  In this
respect we have found  Halley's improvement \cite{Hal}
to the Newton-Raphson
algorithm to be useful.  After $i$ iterations, this leads to the
estimate for the root $y$ of a function $f(y)$ as
\be
y_{i+1} \E y_i - f(y_i) \, \left [ \, f'(y_i) - \frac{f(y_i)\,f''(y_i)}{2  
f'(y_i)} \, \right ]^{-1} \> ,
\ee
which, with $f(y) = x - y \tan y$ and $x$ fixed, yields
\bea
   y_{i+1} \EA
   y_i \, + \left ( x-y_i \tan y_i \right ) \, \Biggl [ \,  
              y_i (1+\tan^2 y_i) + \tan y_i \nonumber \\
 && \hspace{2cm} + \, 
               \frac{\displaystyle (y_i \tan y_i + 1)(\tan^2 y_i + 1) }
                    {\displaystyle y_i ( 1 + \tan^2 y_i ) + \tan y_i }
               (x-y_i \tan y_i ) \, \Biggr ]^{-1} \> .
\label{wtH}
\eea

\subsection{A variational principle for \wt($x$)}
The quantum mechanical model of Section~\ref{sec:QM} can be used to 
obtain some insight and some approximations
for the \wt-function. It is slightly nonstandard because of the 
energy-dependence 
of the additional interaction but this energy-dependence is linear so that 
it can be easily handled. The Schr\"odinger equation may be cast into the form
\be
H_0 \, \psi(\xi) \E E N(\xi) \, \psi(\xi)
\label{general eigen}
\ee
with $H_0 = -\hbar^2 d^2/(2m \, d\xi^2)$ and $ N(\xi) = 1 + \lambda \, 
\delta(\xi-a/2) $. For a positive definite operator $N$ we can define  
\be
H \E N^{-1/2} H_0 N^{-1/2} \> \> \> , \> \phi \E N^{1/2} \psi 
\label{transform}
\ee
and end up again with a normal Schr\"odinger equation $ H \phi = E \phi $
for which all standard methods of quantum mechanics can be applied. In view of
the finite convergence radius of the series expansion the variational method 
is of particular interest here as it is a non-perturbative method.  
It gives an upper bound for the true energy
\be
E \E \frac{\hbar^2 k^2}{2m} \> \> \le \> \> 
\frac{< \phi_t \, | H | \, \phi_t >}{< \phi_t \, | \, \phi_t>} \> , 
\ee
where $ \phi_t $ is a trial state containing variational parameters. Using 
Eq. (\ref{transform}) this can be cast into the form
\be
E \> \> \le \> \> \frac{< \psi_t \, | H_0 | \, \psi_t >}{< \psi_t \, | N | 
\, \psi_t >} \> , 
\ee
which is a well-known variational principle for the generalized eigenvalue 
problem (\ref{general eigen}) \cite{MoFe}. Recalling
 Eq. (\ref{wavenumber by wt}) and 
setting $ a = \pi$ for convenience we obtain
\be
\wt^2(x) \> \le \> \frac{\pi^2}{4} \,  \int_0^{\pi} d\xi \, \left (
\frac{d \psi_t(\xi)}{d\xi} \right )^2 \> \cdot  \left [ \, \int_0^{\pi} d\xi 
\> \psi_t^2(\xi) 
+ \frac{\pi}{x} \, \psi_t^2\left ( \frac{\pi}{2} \right) \, \right ]^{-1} 
\> , \psi_t(0) = \psi_t(\pi) \E 0 \> .
\label{var}
\ee
As usual this variational principle can be applied most easily to the ground 
state, 
i.e. to the principle branch. Indeed, taking as trial wave function just the 
unperturbed ground state wave function $\> \psi_t(\xi) = \sin \xi \> $ gives
\be
\Wt{1}(x) \> \le \> \frac{\pi}{2} \sqrt{ \frac{x}{x+2}} \> \> 
 , \hspace{0.5cm} \left ( \frac{1}{x} \> > \> - \frac{1}{2} \right )
\label{var approx1}
\ee
which covers both limiting cases of the series expansion studied in the 
previous Section~\footnote{Note that $\pi/(2 \sqrt{2}) = 1.1107$ so that 
the bound 
is also fulfilled near $x = 0$.}. However, it predicts a square-root type 
branchpoint on the real axis at  $x = - 2$ (where $ < \psi_t \, | N | \,  
\psi_t > $ vanishes) which obviously is not realistic.
One can easily do better by allowing for an admixture of the
first excited symmetric state, i.e. 
$ \> \psi_t(\xi) = \sin \xi + b \sin 3 \xi \> $, where $b$ may be optimized 
for each value of $x$. A simple calculation in which we
fix the square root sign by demanding that $b \to 0$ for $x \to \pm \infty$ 
then gives
\be
\Wt{1}(x) \> \le \> \frac{3\pi}{2} \left ( \frac{x}{ 5x + 10 + 2 \, 
{\rm sgn}(x) \sqrt{25+16x+4x^2}} \right )^{1/2} \>  \> .
\label{var approx2}
\ee
This improves the bound near $ x = +0 $ to 
$ \> \> \Wt{1}(x) \le 3 \pi \sqrt{5 x}/20 = 1.0537 \, \sqrt{x} \> $ 
and predicts $ \Wt{1} (-0) \le \pi \sqrt{5}/2 = 1.1180 \, \pi $. 

\unitlength1mm
\begin{figure}[htb]
\bce
\mbox{\epsfysize=10cm\epsffile{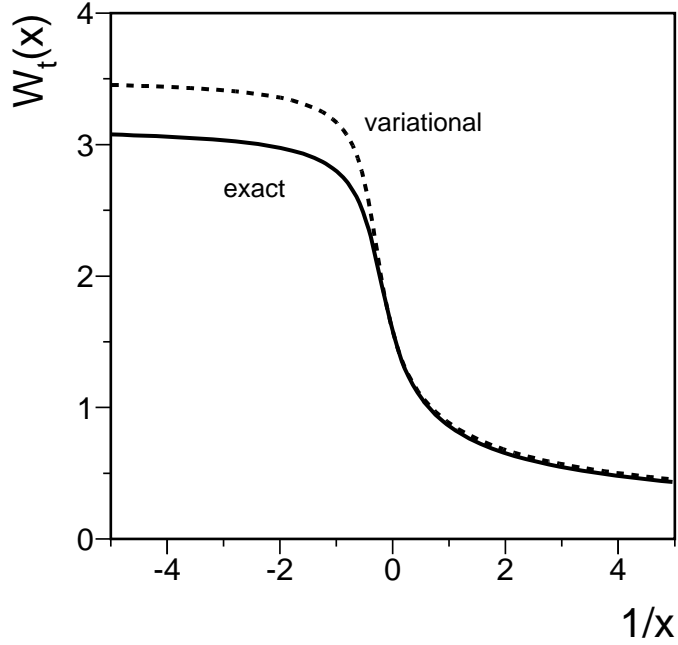}}  
\vspace*{-1.2cm}
\ece
\caption{The principal branch of the \wt-function as function of $1/x$, as 
well as its 
variational approximation of Eq.~(\protect{\ref{var approx2}}).}
\label{fig:var}
\vspace{0.5cm}
\end{figure} 

Fig. \ref{fig:var} compares this variational approximation 
with the exact $\Wt{1}$-function calculated numerically via the iteration 
procedure.
For $x > 0$ the agreement is excellent.
While the approximation is regular on the negative $x$-axis it
exhibits branchpoints at $ x = -2 \pm 3 i/2 $ in the complex $x$-plane where 
the inner square root vanishes. This would restrict 
the series expansions of $\Wt{1}(x)$ around $x = 0$ or $x = \infty$ to 
converge for 
magnitudes smaller or larger than $5/2$, respectively. In the next Section 
we will find that the exact convergence radius is $2.6397$ .

\vspace{1cm}

\section{Analytic structure in the complex plane} 
\label{SecAStr}

\setcounter{equation}{0}

\subsection{The branch points}

The implicit equation (\ref{Wt}) can be written in the form
\be
    y \E {\rm Arctan} \frac{x}{y} \E  \frac{1}{2i} \, {\rm Ln} \, 
{\frac{y+ix}{y-ix}}
\label{yatanxy}
\ee
where $ {\rm Arctan}, \> {\rm Ln} $ denote the general (multivalued) 
inverse tangent and logarithmic functions, respectively.
One can immediately see that the function $y=\wt(x)$
is finite in any finite part of the complex $x$ plane.  Indeed,
the only singularities of the expression on the r.h.s. of Eq. (\ref{yatanxy}),
as a function of $x$ and $y$, are the logarithmic branch points
at $y=\pm ix$. The latter cannot be satisfied simultaneously with 
Eq. (\ref{yatanxy}) at any finite $x$.
  The derivatives of $y=\wt(x)$ have a rather simple structure (in implicit 
form):
\bea
    y' \EA \frac{dy}{dx} \E \frac{y}{y^2+x^2+x} =
                                \frac{\cos^2 y}{y+\sin y \cos y }
\label{y1}
\\
    y'' \EA \frac{d^2y}{dx^2} \E
                 - \frac{2xy}{(y^2+x^2+x)^2} - \frac{2y^3}{(y^2+x^2+x)^3}
\label{y2}
\\
    y^{(n)} \EA \frac{d^n y}{dx^n} \E \frac{P_n(x,y)}{(y^2+x^2+x)^{2n-1}}
\label{yn}
\eea
where $P_n(x,y)$ is some polynomial in $x$ and $y$. The first derivative 
diverges at the points where
\be
     y^2 + x^2 + x \E 0
\label{divder}
\ee
together with the all higher derivatives.  Because all the derivatives have
the same polynomial $(y^2+x^2+x)$ to an appropriate power in the denominator,
they are either all finite or all divergent at the same point $x$.
Therefore, in order to find all singularities of $\wt(x)$ it
is sufficient to solve Eq. (\ref{divder}); this is equivalent to
\be
    \sin y  \, \cos y + y \E 0
\label{yeq}
\ee
which involves only the function value $y$.
One obvious solution is $y_0 = 0$; all other solutions of
(\ref{yeq}) are complex. Setting $2y = u + i v$, where $u$ and $v$ are
real, gives
\bea
    u + \sin u \,  \cosh v \EA 0
\label{equ}
\\
    v + \cos u \, \sinh v \EA 0 \> .
\label{eqv}
\eea
We can restrict ourselves to $u > 0$ (if $y$ is a solution of (\ref{yeq})
then $-y$ is also a solution).
Because $\cosh{v}$ and $\sinh{v}/v$ are always positive, we get
the following constraints from Eqs. (\ref{equ}, \ref{eqv})
\be
  \sin u \> \leq  \>  0 \> , \quad
  \cos u  \> \leq \>  0 \> ,
\ee
which leads to
\be
   ( 2 n - 1 ) \, \pi \leq u \leq \left ( 2 n - \frac{1}{2} \right ) \, \pi, 
\quad n \E 1 , 2, \ldots
\label{uint}
\ee
From Eqs. (\ref{equ}, \ref{eqv}) one gets an equation containing only 
the variable 
$u$:
\be
   \tan u \, \cdot \, {\rm arccosh} \left (-\frac{u}{\sin u} \right )
   \E \sqrt{u^2 - \sin^2 u}
\label{equo}
\ee
which has exactly one solution $u_n$ in each interval $n$ defined by
Eq. (\ref{uint}). For large $n$ one finds 
$ u_n \E  b_n - \ln \left ( 2 b_n\right )/b_n  + \ldots $ 
where $  b_n = ( 2n - 1/2 ) \, \pi $ and therefore
\bea
\pm \, y_n \EA  \frac{1}{2} \, b_n \pm \frac{i}{2} \ln \left ( \, 2 b_n \, 
\right ) + {\cal O} \left ( \frac{\ln n}{n} \right )\\
x_n \EA  -\frac{1}{2} \ln \left (\, 2 b_n \, \right ) - \frac{1}{2} \, \pm \,
 \frac{i}{2} \, b_n \, + {\cal O} \left ( \frac{\ln n}{n} \right ) 
\eea
which is in good agreement with the numbers in Table~\ref{tab:branch points}
obtained from the numerical solution of Eq. (\ref{equo}).

\begin{table}[htb]
\bce
\begin{tabular}{|c|c|c|c|}
\hline
      &                              &              &                              \\
$n$   &  $x_n$                       & $|x_n|$      &     $\pm \,y_n$              \\
      &                              &              &                              \\
\hline
      &                              &              &                              \\
 1    & $-1.650611 \pm 2.059981 \,i$ &   2.639705   & $2.106196 \pm 1.125364 \,i$  \\
 2    & $-2.057845 \pm 5.334708 \,i$ &   5.717853   & $5.356269 \pm 1.551574 \,i$  \\
 3    & $-2.278470 \pm 8.522637 \,i$ &   8.821948   & $8.536682 \pm 1.775544 \,i$  \\
 4    & $-2.431122 \pm 11.68877 \,i$ &  11.938917   & $11.69918 \pm 1.929404 \,i$  \\
 5    & $-2.547991 \pm 14.84580 \,i$ &  15.062869   & $14.85406 \pm 2.046852 \,i$  \\
 6    & $-2.642706 \pm 17.99809 \,i$ &  18.191069   & $18.00493 \pm 2.141891 \,i$  \\
 \ldots &                          &              &                                \\
\hline        
\end{tabular}
\ece
\caption{The branch points $x_n$ of $y = \wt (x)$ and the corresponding
function values $y_n$. }
\label{tab:branch points}
\end{table}

\vspace{0.5cm}
\noindent
To investigate the type of the singularities
we introduce new variables $\Delta x$ and $\Delta y$
in the vicinity of the singular point $x_n$:
\be
    x \E x_n + \Delta x \> , \quad
    y \E y_n + \Delta y   \> .
\ee
The implicit equation (\ref{Wt}) can now be rewritten in the form
\be
   \Delta x \cos \left ( y_n + \Delta y \right ) \E
   \left ( \, \Delta y \cos \Delta y - \sin \Delta y \, \right ) \, \sin y_n
+ \Delta y \, \sin \Delta y \; \cos y_n \> .
\label{eqdxdy}
\ee
If we go along a closed contour around the point $y_n$ in
the complex $y$ plane assuming that $\Delta y$ is sufficiently small,
then the corresponding path in the complex $x$ plane will make a closed 
contour
winding up twice as many times around the corresponding point $x_n$.
This can be demonstrated by inspecting the leading term in the expansion
of Eq. (\ref{eqdxdy}) in powers of $\Delta y$:
\be
   \Delta x \cos{y_n} \E (\Delta y)^2 \cos{y_n} + {\cal O}((\Delta y)^3)
\ee
Therefore all singularities of $\wt(x)$ are of square root type : 
$\Delta y = \pm (\Delta x)^{1/2}+\ldots$ or
\be
   \wt(x)  \> \stackrel{x\to x_n}{\longrightarrow} \> \wt(x_n) \pm 
\sqrt{x-x_n} \> .
\label{wtxbp}
\ee
Another way to obtain this result is to insert the following ansatz
\be
   \wt(x) \> \stackrel{x\to x_n}{\longrightarrow} \>   \wt(x_n) + c \, 
\left ( x - x_n \right )^{\kappa}
\label{ansatz}
\ee
into the differential equation (\ref{diff eq}).   Here $\kappa $ must be 
positive since we know that $\wt(x_n)$ exists. Assuming $\kappa < 1 $ 
we obtain in leading order
\be
\kappa  c  (x-x_n)^{\kappa - 1} \E  \frac{1}{2 c} (x-x_n)^{-\kappa}
\ee
which determines $\kappa = 1/2$ and $c^2 = 1$.

\subsection{The Riemann surface}

Since the function $\wt(x)$ has an infinite (countable) number of
branch points, it has an infinite number of Riemann sheets as was
already clear from Fig. \ref{fig:branches}. Therefore the question
arises how to define these Riemann sheets in the most convenient way
for any particular purpose.  As we know the position and the type of
all branch points, it only remains to define the cuts originating from
these branch points.  Given the cuts (we call any complete set of them
a {\it scheme}), one then only needs to know how the different Riemann
sheets are connected across the cuts.

It turns out that the labeling of the Riemann sheets based on the inspection 
of the function on the real $x$-axis without a proper consideration of the 
branch points, as was done in Section~\ref{sec:QM}, is not very illuminating 
when the whole complex $x$-plane is concerned.  Below we consider two other 
schemes of Riemann sheets which are better suited for the discussion of the 
singularities and analytical continuation across the cuts.  In order to 
keep notations simple 
we prefer not to indicate the particular scheme explicitly
in the notation $\Wt{n}(x)$; it will be defined in the context whenever the
scheme is relevant. 

\begin{figure}
\bce
\mbox{(a) \hspace*{75mm} (b) \hspace*{40mm} }\\[-\baselineskip]
\mbox{\epsfxsize=150mm \epsffile{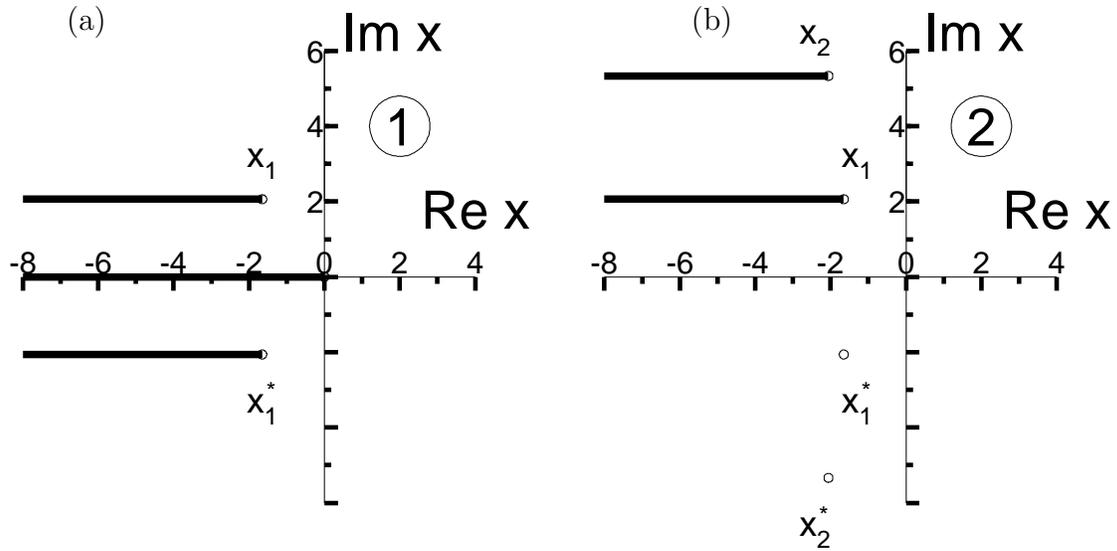}}
\ece
\caption{\label{fig:RSinf}%
The Riemann sheets of $y = \wt(x)$ in the scheme where all cuts 
go to $-\infty$: (a) sheet 1, (b) sheet 2.}
\vspace{0.5cm}
\end{figure}

The first new scheme of Riemann sheets is demonstrated in Fig. \ref{fig:RSinf}
where the cuts originating from the branch points $x_n$ run to $-\infty$ along
the lines of constant imaginary part $\mbox{Im}\; x = \mbox{Im}\; x_n$,
$-\infty < \mbox{Re}\; x \leq \mbox{Re}\; x_n$.  
%
%
Every Riemann sheet can be uniquely specified by the limit values
at ${x\to+\infty}$ or ${x\to +0}$ as given in Table \ref{tab:cuts to infinity}.

\vspace{0.5cm}

\begin{table}[htb]
\bce
\begin{tabular}{l|cccccccc}
\hline
      &        &    &   &   &   &        &     &                 \\
sheet & \ldots & -1 & 1 & 2 & 3 & \ldots & $n$ & \ldots          \\
$\lim_{x\to+\infty} \Wt{n}(x)$ &
  \ldots & $-\pi/2$ &
            $\pi/2$ & $3\pi/2$ & $5\pi/2$ & \ldots &
            $\mbox{sgn}(n)(|n|-1/2)\pi$ & \ldots               \\
$\lim_{x\to +0} \Wt{n}(x)$ &
  \ldots &  $-0$ &
            $+0$ & $\pi$ & $2\pi$ & \ldots & $\pi (n-1)$ & \ldots \\
      &        &    &   &   &     &        &             &         \\
\hline
\end{tabular}
\ece
\caption{The scheme of Riemann sheets where the cuts extend to $-\infty$ 
as shown in Fig.~\protect\ref{fig:RSinf}.}
\label{tab:cuts to infinity}
\end{table}

\vspace{0.5cm}
\noindent
There is a finite number of cuts on every sheet.
The connection between different sheets is demonstrated in Fig. 
\ref{fig:RSinf}. On all sheets $n$ except $ n = \pm 1 $, the function 
$\wt(x)$ is continuous and
real on the real $x$ axis. On the sheets $n = \pm 1$, the function is real for
$ x \geq 0 $ and purely imaginary for $ x<0 $.

\begin{figure}
\bce
\mbox{(a) \hspace*{80mm} (b) \hspace*{40mm} }\\[-\baselineskip]
\mbox{
\mbox{\epsfxsize=80mm \epsffile{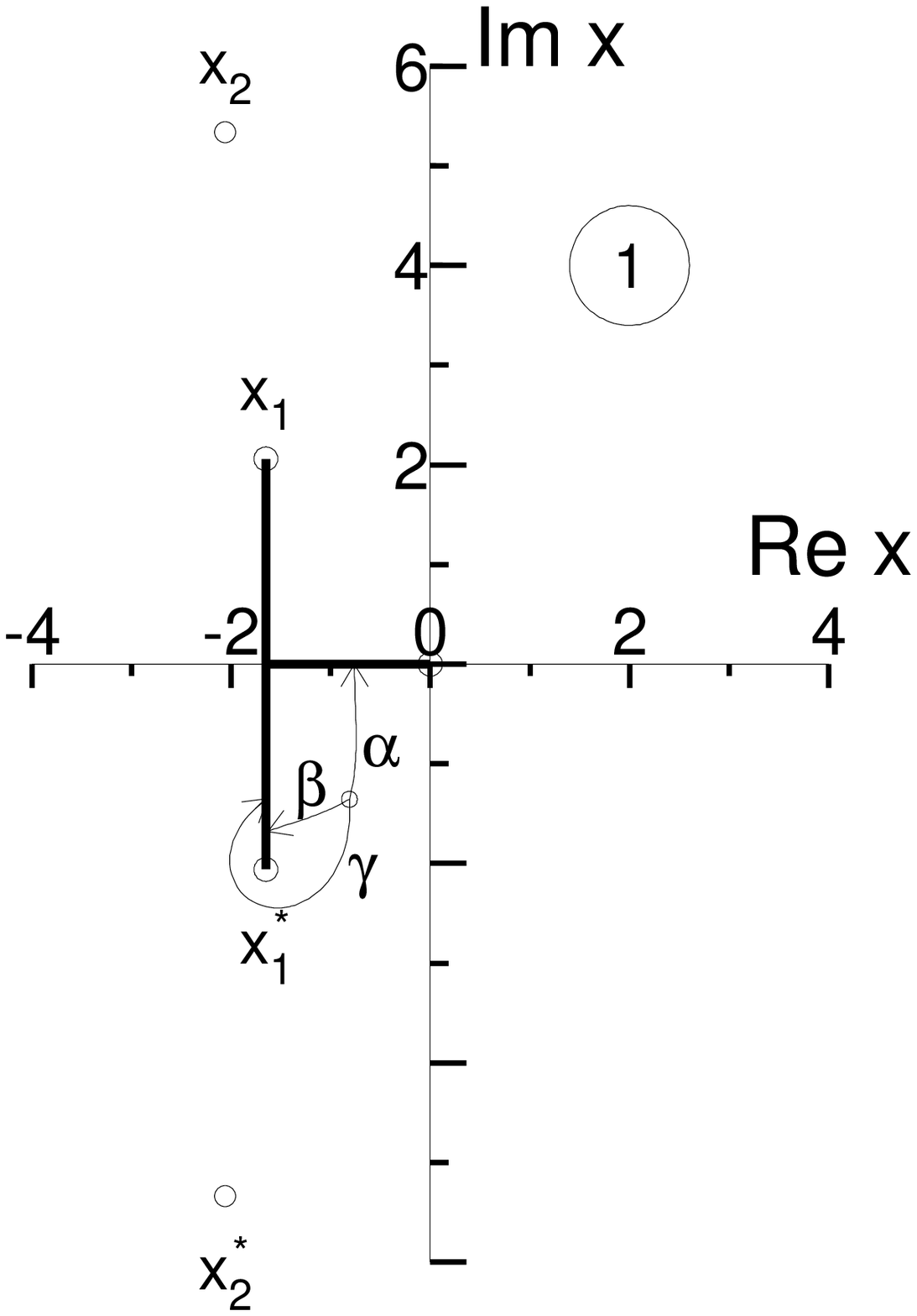}} 
\mbox{\epsfxsize=75mm \epsffile{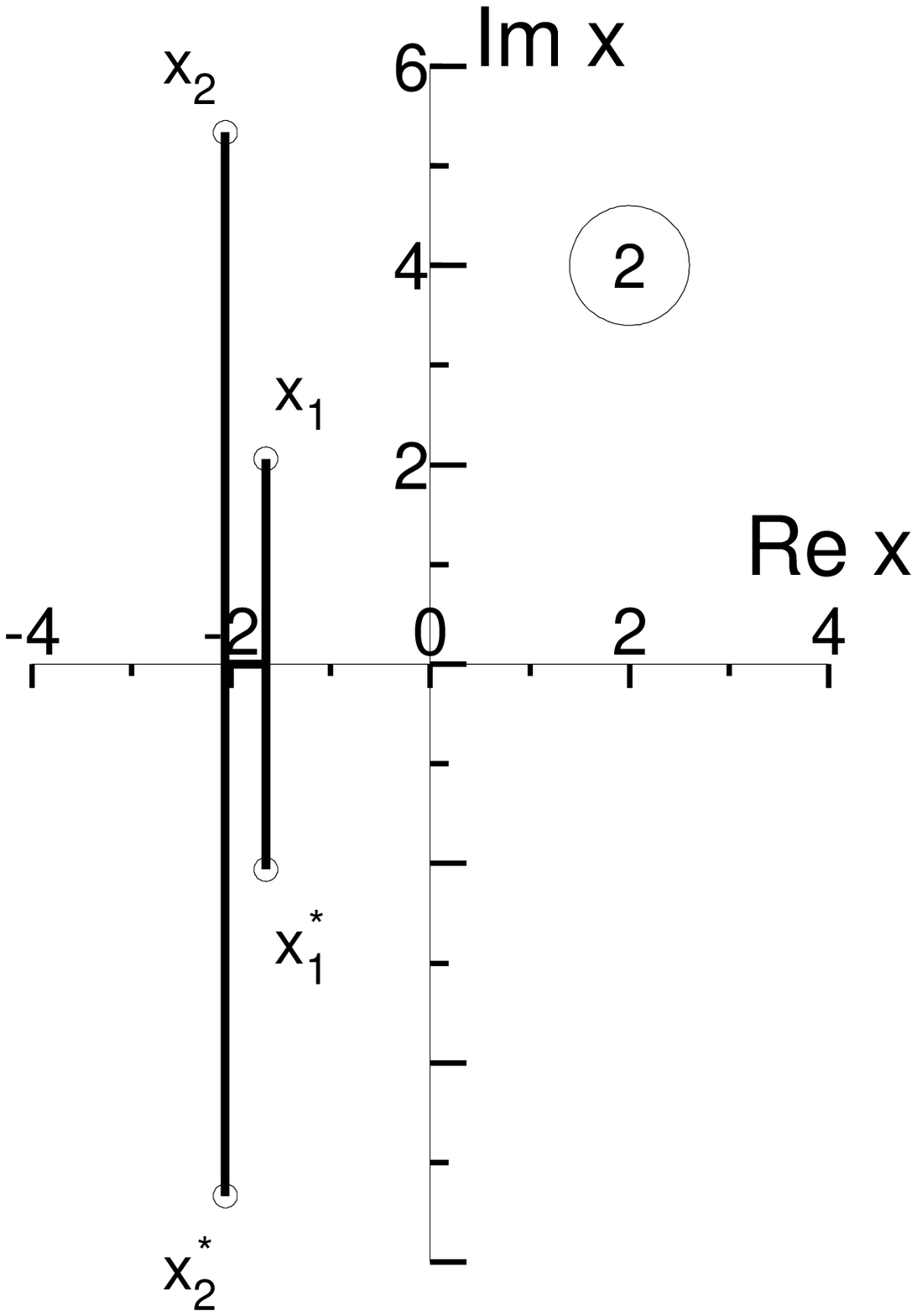}}
}
\\
\mbox{(c) \hspace*{40mm} }\\[-2\baselineskip]
\mbox{\epsfxsize=150mm \epsffile{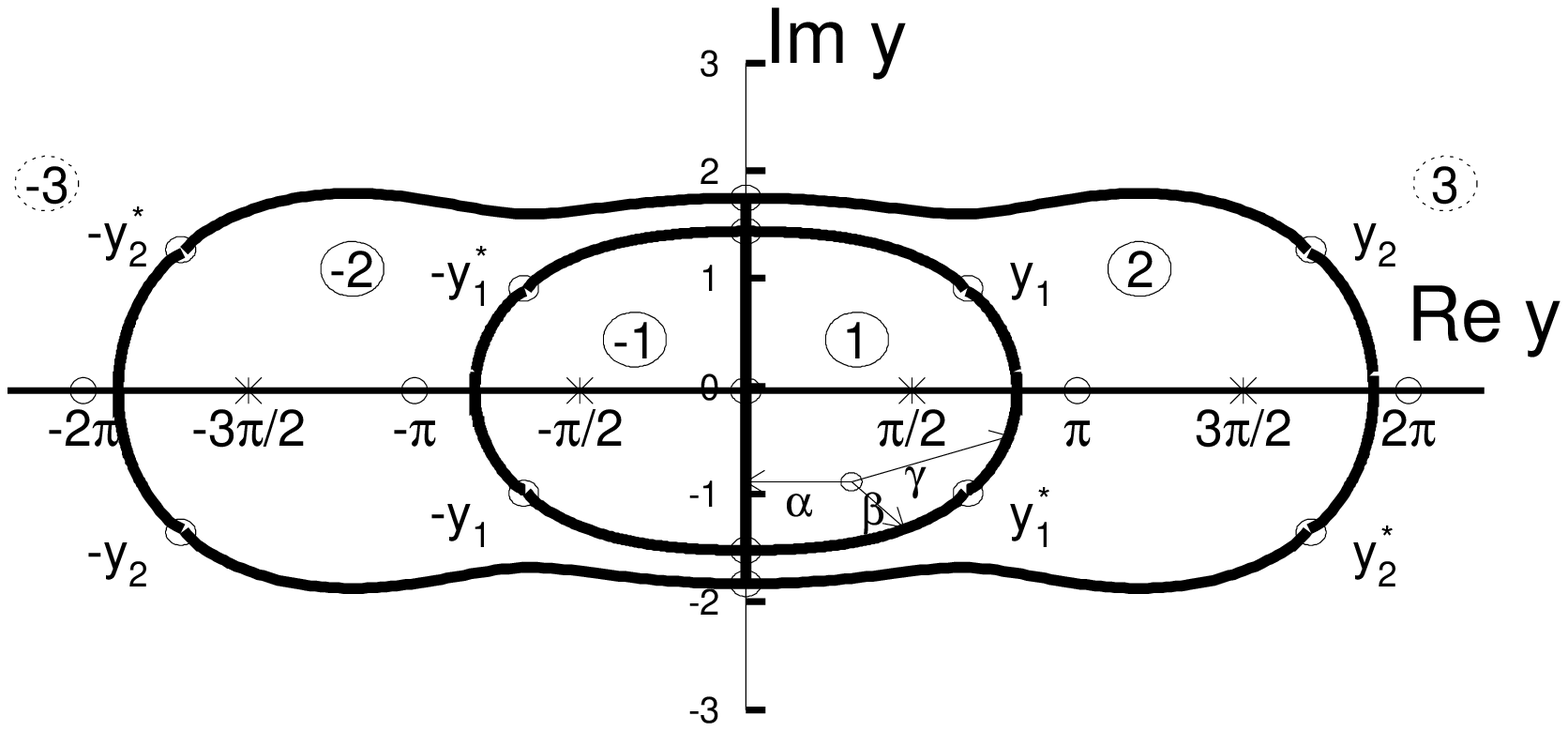}}
\ece
\caption{\label{fig:RS}%
The Riemann surface of $y = \wt(x)$: (a) sheet 1, (b) sheet 2, and 
(c) the images of different sheets in the complex $y$ plane.
The path $\alpha$ in plot (a) goes from sheet $1$ to sheet $-1$ across the cut
$(\mbox{Re}\;x_1, 0)$, the corresponding image of this path in the $y$ plane
is shown in plot (c).
The paths $\beta$ and $\gamma$ go from sheet $1$ to sheet $2$ across
the cuts $(\mbox{Re}\;x_1,x_1^*)$ and $(x_1, x_1^*)$ respectively.
}
\vspace{0.5cm}
\end{figure}

\begin{figure}
\bce
\mbox{\epsfysize=70mm \epsffile{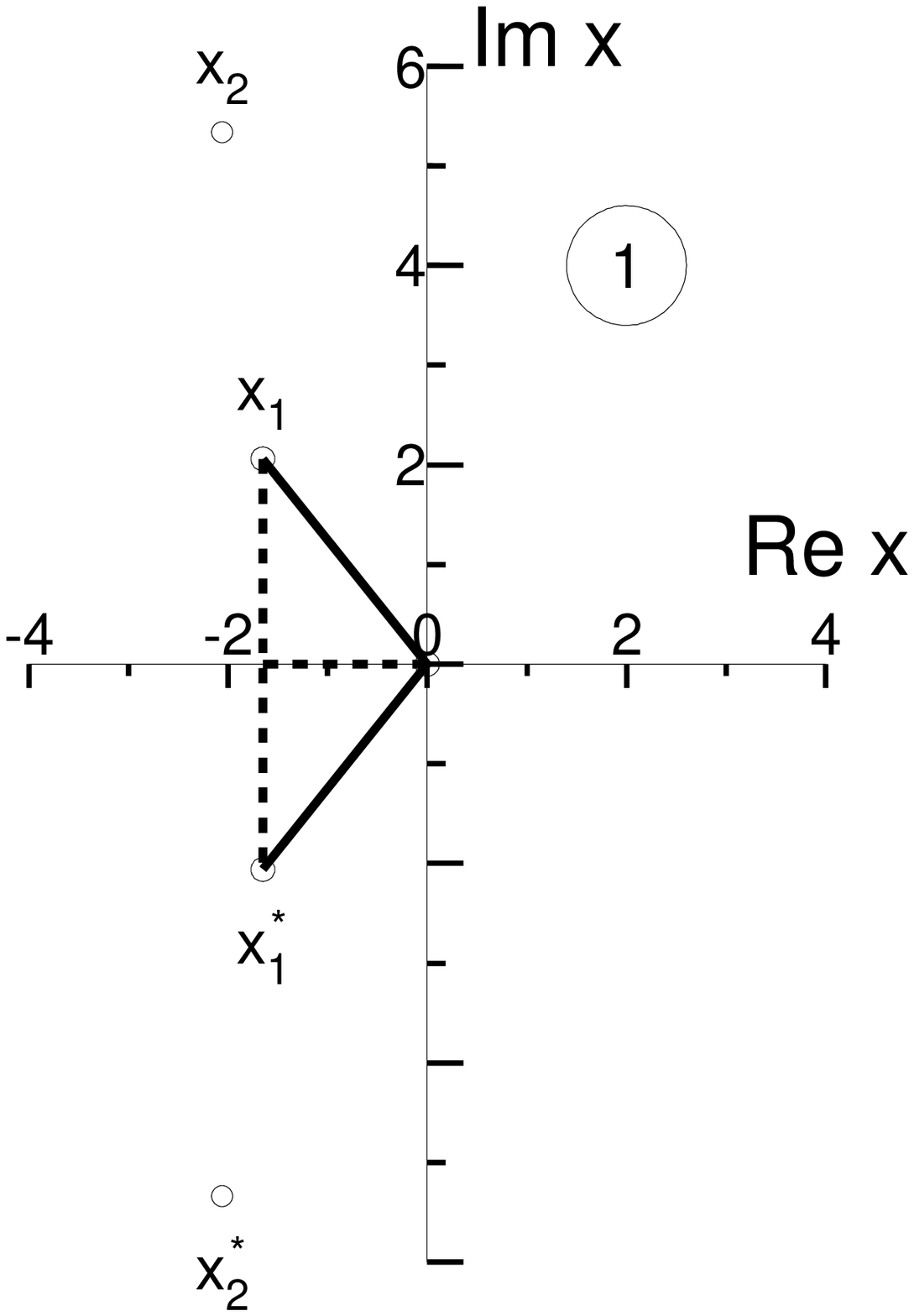}}
\ece
\caption{\label{fig:RSmod}%
A modification of the main Riemann sheet of $y = \wt(x)$ shown in 
Fig.\protect\ref{fig:RS} (a): 
the cut between $x_1$ and $x_1^*$  is deformed 
so that it crosses the real $x$-axis by going over the branch point $x=0$. }
\end{figure}

Sending all cuts to infinity has, however, a drawback: it obscures
the analytical properties of $\wt(x)$ at $ x = \infty $.  In fact, the points
$ x = \infty $ correspond to the regular points $ t = 1/x = 0 $ of the 
function $\wt(1/t)$.  
In the case when the analytical properties at infinity are of interest, 
alternative schemes of the Riemann surface, such as shown in Fig. 
\ref{fig:RS}, have definite advantages.   
This second new scheme provides a convention of branches which is very 
suitable for the quantum mechanical example in Section \ref{sec:QM}, 
where the argument $x$ was proportional to the inverse coupling constant 
so that the perturbative expansion 
became an expansion of $\wt$ around $ x = \infty $.  
In this scheme all cuts are finite in the complex $x$-plane:
every branch point $x_n$ ($ n = 1, 2, \ldots $ ) is connected with the 
corresponding complex conjugated point $x_n^*$, and every sheet, except 
for the sheets $\pm 1$, has a pair of such cuts.
By deforming the cut connecting the branch points $x_1$ and $x_1^*$ as shown  
in Fig.~\ref{fig:RSmod} one can make it cross the real $x$-axis at $x=0-$, 
so that the function is made continuous at both $x>0$ and $x<0$, as in the 
case considered in Section~\ref{sec:QM}.  However, this comes at the price of 
having the branch point at $x=0$ superimposed on the cut between $x_1$ and 
$x_1^*$, which is 
not very convenient for the purpose of analytical continuation.

The cut originating at $x_0 = 0$ occurs on every sheet in a finite interval:
on the sheets $\pm 1 $ it goes along the real axis from zero to the cut
connecting $x_1$ and $x_1^*$, on all other sheets it lies on the real axis
between the two cuts connecting $x_n$ with $x_n^*$ and $x_{n-1}$ with 
$x_{n-1}^*$.
For every sheet $n$ all the cuts are located within the limits
$|x|\leq |x_n|$, $\mbox{Re}\;x_{n} \leq \mbox{Re}\;x \leq \mbox{Re}\;x_{n-1}$.
The Riemann sheets in this scheme can be uniquely specified by the
function values at infinity, which is a regular point on every sheet 
(see Table \ref{tab:cuts finite}).

\vspace{0.2cm}
\begin{table}[htb]
\bce
\begin{tabular}{l|cccccccc}
\hline
      &        &    &   &   &   &        &   &        \\
sheet & \ldots & -1 & 1 & 2 & 3 & \ldots & n & \ldots \\
$\lim_{x\to\infty} \Wt{n}(x)$ &
  \ldots & $-\pi/2$ &
            $\pi/2$ & $3\pi/2$ & $5\pi/2$ & \ldots &
            $\mbox{sgn}(n)(|n|-1/2)\pi$ & \ldots      \\
      &        &    &   &   &   &        &   &        \\
\hline
\end{tabular}
\ece
\caption{The scheme of Riemann sheets with the finite cuts shown 
in Fig.~\protect\ref{fig:RS}.}
\label{tab:cuts finite}
\vspace{0.5cm}
\end{table}

\noindent
This scheme will be used below in all references to different
branches of $\Wt{n}(x)$.
The sheets $n$ and $-n$ differ only by the function sign:
if $ y = \Wt{n}(x)$, then $ -y = \Wt{-n}(x)$ (see Eq.~(\ref{negative n})).
The connection between different sheets is shown in Fig. \ref{fig:RS}.

The simplest way to investigate this connection is to begin with
some path in the $y$ plane (image) that connects two points on different
sheets for the same $x$ and to calculate
the corresponding path in the $x$ plane using Eq. (\ref{Wt}).
 For any $n = \pm 1, \pm 2, \ldots$, the sheets $n$ and $-n$ are connected
to each other across the cut on the real axis between the points
$\mbox{Re}\;x_n$ and $\mbox{Re}\;x_{-n}$
(see the path $\alpha$ in Figs. \ref{fig:RS} (a,c)).
The cut between the branch points $x_n$ and $x_{n}^*$ connects
the sheet $n$ with the sheet $(n+1)$ for $n>0$ or with the sheet
$(n-1)$ for $n<0$
(the paths $\beta$ and $\gamma$ in Figs. \ref{fig:RS} (a,c) ).
The structure of the cuts for all sheets with $|n|>2$ is similar
to that of sheet 2 (see Fig. \ref{fig:RS} (b) ).

The definition of $\wt(x)$ as an implicit function, Eq. (\ref{Wt}),
involves the function $ y\tan y $ which is {\it single valued} on the
compactified complex plane. Therefore for any $y$ there is {\it one and only
one} value of $x$ such that $\wt(x) = y$. This allows us to determine the
topology of any part of the Riemann surface by inspecting its projection
onto the $y$ plane.
The topological structure of the Riemann surface of $\wt(x)$ turns out
to be quite simple: any set of connected Riemann sheets
with $n = \pm 1, \pm 2, \ldots, \pm n_{max}$ is homeomorphic to
a two-dimensional disk as illustrated in Fig. \ref{fig:RS} (c).
For example, the part of the Riemann surface that consists of four sheets
with $n = \pm 1, \pm 2$ is a manifold with a boundary homeomorphic to the
one-dimensional sphere $S^1$.
The Euler characteristic \footnote{This is the topological 
invariant given by $\chi = V-L+S $ where $L$, $V$, and $S$ are the number 
of vertices, links and faces respectively in a triangulation of the manifold 
(see e.g. Ref. \cite{Mas}). For example,
the triangulation with the vertices corresponding to the branch points
and the links corresponding to the cuts in Figs. \ref{fig:RS} (a,b)
has $V = 13$, $L = 16$, $S = 4$.} of this manifold is $\chi = 1$. This is equal to
the Euler characteristic of a disk with a boundary.
If this manifold is extended by adding
pairs of sheets connected to its boundary ($n = \pm 3, \pm 4, \ldots$) then
the Euler characteristic remains unchanged.
While much more could be said about the topological structure of the 
Riemann surface a further discussion is beyond the scope of this paper.
We close by pointing out that, in a certain sense, the Riemann surface of 
$\wt(x)$ resembles an elaborate {\it origami}
where a square of paper (which is topologically equivalent to a
two-dimensional disk) is folded \footnote{The origami paper is not stretched, 
so the analogy ceases to be valid  when we want
to deform the disk into the Riemann surface under consideration.}
into a beautiful shape {\it without paper cuts}.  
The same property holds true for every function defined by an implicit 
equation involving any functions that are single--valued at finite $x$. 

\begin{figure}[htb]
\bce
\mbox{(a) \hspace*{75mm} (b) }\\[-\baselineskip]
\mbox{\epsfxsize=150mm \epsffile{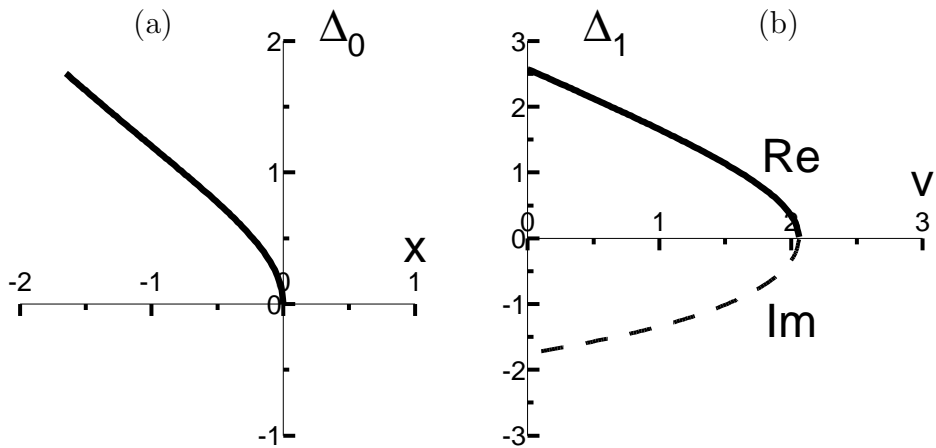}}
\ece
\caption{\label{fig:Disc}%
(a) The imaginary part of the discontinuity of $\Wt{1}(x)$ across the cut
    $-\infty < x \leq 0$,
(b) the discontinuity of $\Wt{1}(x)$ across the cut connecting $x_1$ and 
$x_1^*$.
}
\vspace{1cm}
\end{figure}

\subsection{The dispersion relation for $\wt(x)$}

The analytic structure discussed above allows to 
derive the dispersion relation satisfied by
$\wt(x)$ on the first sheet 
\bea
   \Wt{1}(z) \EA \frac{\pi}{2} +
   \frac{1}{\pi} \int_a^0 du \, \frac{\Delta_0(u)}{u-z}   +
   \frac{1}{\pi} \int_0^b dv \, \left(
         \frac{\Delta_1(v)}{a+iv-z} + \frac{\Delta_1^*(v)}{a-iv-z} \right) 
\> , \\
a \EA \mbox{Re}\; x_1 \> , \quad  b \E \mbox{Im}\; x_1 \> .
\eea
Here $\Delta_0(u)$ and $\Delta_1(v)$ are determined by the discontinuities
across the cuts on sheet 1 (see Fig. \ref{fig:Disc}):
\bea
        \Delta_0(u) \EA \frac{1}{2i} \left [ \Wt{1}(u+i\epsilon) - \Wt{1}
(u-i\epsilon) \right ]
        = \mbox{Im}\; \Wt{1}(u+i\epsilon) \, ,
        \, -\infty < u \leq 0
\\
    \Delta_1(v) \EA \frac{1}{2}  \left [ \, \Wt{1}(a+\epsilon+iv) - \Wt{1}
(a-\epsilon+iv)  \, \right ]
        \quad,\quad 0 \leq v \leq b \> .
\eea

\vspace{2cm}

\section{Concluding remarks}

In the preceding Sections we have derived a number of results on the 
analytic structure, the numerical evaluation, series expansions
and a simple quantum mechanical problem where the function $\wt(x)$ 
makes its appearance. While certainly not being exhaustive, these results 
demonstrate the amazingly rich structure and the fascinating properties of 
the \wt-function which is defined by a simple, implicit equation. We hope  
that our selected findings may be useful for other researchers encountering
this function. Of course, the methods used may be also adapted
to related functions and indeed one might also consider, 
at the expense of convenience, more general
implicit equations of which Eq. (\ref{Wt}) is just a special case.
For example, Wright has elaborated~\cite{Wright1}
on  equations of the form 
\be
(z + b) \, e^{z+a} \E z - b  
\ee 
and discussed the real solutions $z$ to these equations 
as a function of $a$ and $b$ in Ref.~\cite{Wright2}.
Eq. (\ref{Wt}) may be
brought into this form, with $ a = 0 $ and $ b = 2 \, x $, by writing the
trigonometric function in its exponential form and setting 
$ \wt = i z/2 $. Our results in Section 2 (and elsewhere) would then
correspond to purely imaginary, special solutions of Wright's equation.

\newpage


\end{document}